\input harvmac
\Title{\vbox{
\hbox{HUTP-00/A016}
\hbox{\tt hep-th/0005180}}}{Unifying Themes in Topological
Field Theories}
\bigskip
\centerline{Cumrun Vafa}
\bigskip
\centerline{Jefferson Physical Laboratory}
\centerline{Harvard University}
\centerline{Cambridge, MA 02138, USA}
\vskip .3in
We discuss unifying features of topological
field theories in 2, 3 and 4 dimensions.  This includes
relations among enumerative geometry
(2d topological field theory) link invariants (3d Chern-Simons
theory) and Donaldson invariants (4d topological theory).
(Talk presented in conference
on Geometry and Topology in honor of M. Atiyah, R. Bott,
F. Hirzebruch and I. Singer, Harvard University, May 1999).

\Date{May 2000}

\newsec{Introduction}
There has been many exciting interactions between physics
and mathematics in the past few decades.  Many of these
developments on the physics side are captured by
certain field theories, known as topological field theories.
The correlation function of these theories compute
 certain mathematical invariants.  
Even though the original motivation
for introducing topological field theories
was to gain insight into
these mathematical invariants, topological field theories have been found
to be important for answers to many questions of interest
in physics as well.

The aim of my talk here is to explain certain connections
that have been discovered more recently
 among various topological field theories.
I will first briefly review what each one is, and then go on
to explain some of the connections which has been discovered between them.

The main examples of topological field theories that have been proposed
appear in dimension two \ref\wittt{E. Witten, ``Topological
Sigma Model," Comm. Math. Phys.  {\bf 118} (1988) 411.}\
known as topological sigma models, in dimension three \ref\witcs{E. Witten,
``Quantum Field Theory and the Jones Polynomial,'' Comm. Math. Phys.
{\bf 121} (1989) 351.}\ known as
Chern-Simons theory and in dimension four \ref\wittf{E. Witten, ``Topological
Quantum Field Theory,'' Comm. Math. Phys. {\bf 117} (1988) 353.}\
known as topological Yang-Mills theory.  The 2d and the 4d topological
theory are related to an underlying supersymmetric quantum field theory,
and there is no difference between the topological
and standard version on the flat space. The difference
between conventional supersymmetric theories and topological ones in these
cases only arise when one considers curved spaces.  In such
cases the topological version, is a modified version of the 
supersymmetric theory on flat space where some of the fields have different
Lorentz transformations properties (compared to the conventional
choice). This modification of Lorentz transformation
properties is also known as twisting, and is put in primarily
to preserve supersymmetry on curved space. In particular
this leads to having at least one nilpotent supercharge $Q$ as a 
scalar quantity, as opposed
to a spinor, as would be in the conventional spin assignments.
The physical observables of the topological
theory are elements of the $Q$ cohomology.  The path integral
is localized to field configurations which are annihilated
by $Q$ and this typically leads to some moduli problem
which lead to mathematical invariants.

  In these
theories the energy momentum tensor  is $Q$ trivial,
i.e.,
$$T_{\mu \nu}=\{ Q,\Lambda_{\mu \nu} \}$$
which (modulo potential anomalies) leads
to the statement that the correlation functions
are all independent of the metric on the curved space, thus
leading to the notion of topological field theories (i.e.
metric independence).

The case of the 3d topological theory, is somewhat different.
In this case, namely the case of Chern-Simons theory, one
starts from an action which is manifestly independent of the
metric on the 3 manifold, and thus topological nature
of the field theory is manifest.  

The organization of this paper is as follows:  In section 2
I briefly review each of the three classes of topological theories
and discuss how in each case one goes about computing the correlation
functions.  In section 3 I discuss relations between 2d and
4d topological theories.  In section 4 I discuss relations
between 2d and 3d topological theories.  

\newsec{A Brief Review of Topological Field theories}
In this section I give a rather brief review of topological
field theories in dimensions 2, 3 and 4.

\subsec{TFT in $d=2$: Topological Sigma Models}
Topological sigma models are based on $(2,2)$ supersymmetric
theories in 2 dimensions.  These typically arise
by considering supersymmetric
sigma models on Kahler manifolds.  In other
words, we consider maps from 2 dimensional
Riemann surfaces $\Sigma$ to target spaces $M$ which are Kahler
manifolds (together with fermionic degrees of freedom
on the Riemann surface which map to tangent vectors
on the Kahler manifold).  The topological
theory in this case localizes on holomorpic maps from
Riemann surfaces to the target:
$$X: \qquad \Sigma \rightarrow M $$
$${\overline \partial } X=0$$
If we get a moduli space of such maps we
have to evaluate an appropriate class over it.
  This class is determined by the
topological theory one considers
 (for precise mathematical definitions see
\ref\kco{D.A. Cox and
S. Katz, {\it Mirror Symmetry and 
Algebraic Geometry}, Math Surveys and Monographs, {\bf 68} (ASMS, 1999).}).
Also
 there are two versions of this topological
theory:  coupled or uncoupled to gravity.
Coupling to gravity in this case means allowing
the complex structure of $\Sigma$ to be arbitrary
and looking for holomorphic curves over the entire
moduli space of curves.  The case coupled
to gravity is also sometimes referred to as
`topological strings'.  

A particularly interesting class of sigma modelds
both for the physics as well as for mathematics,
corresponds to choosing $M$ to be a Calabi-Yau
threefold, and considering topological strings
on $M$.  In this case the virtual dimension
of the moduli space of holomorphic maps is zero.
If this space is given by a number of points, the topological
string amplitude just counts how many such points there
are, weighted by $e^{-k(.)}$ where $k(.)$ is the area of the
holomorphic
map (pullback of the Kahler form integrated over the surface)
times $\lambda^{2g-2}$, where $g$ denotes the genus of the
Riemann surface and $\lambda$ denotes the string coupling
constant.  More generally the space of holomorphic
maps will involve a moduli space.  This space comes
equipped with a bundle with the same dimension as
the tangent bundle (the existence of this bundle
and the fact that its dimensions is the same as the tangent
bundle 
follows from the fact that the relevent index is zero).
Topological string computes the top Chern class of such
bundles again weighted by $e^{-k(.)}\lambda^{2g-2}$. These
have to be defined carefully, due to singularities
and issues of compactifications, and lead in general
to rational numbers.
  The sum of these numbers for a given
class $v\in H_2(M,{\bf Z})$ and fixed genus $g$, which we
 will denote by $r_{g,v}$,
 is known as Gromov-Witten invariant.
 We
thus have the full partition function of topological string
 given by
$$F(\lambda ,k)=\sum_{v\in H_2(M,{\bf Z})} r_{g,v} e^{-k(v)}
\lambda^{2g-2}$$
here $k$ denotes the Kahler class of $M$.
Even though the numbers $r_{g,v}$ are not integers, it has been
shown, by physical arguments that $F$ can also be expressed
in terms of other {\it integral} invariants \ref\gopvi{R. Gopakumar 
and C. Vafa, ``M-Theory and Topological Strings I,II,''
hep-th/9809187, hep-th/9812127.}.
These integral invariants are related to certain
aspects of cohomology
classes of moduli of holomorphic curves {\it together
with flat bundles}.

 These invariants associate for each $v\in H_2(M,{\bf Z})$
and eash positive (including 0) integer $s$
a number $N_{v,s}$ which denotes the `net' number of 
BPS membranes with charge in class $v$ and `spin' $s$
(for precise definitions see \gopvi ).  Then we have
\eqn\gova{F(\lambda ,k)=\sum_{n>0,v\in H_2(M,{\bf Z})}{1\over n}
N_{v,s}e^{-n k(v)} [2 Sin(n\lambda /2)]^{2s-2}}
For all cases checked thus far the Gromov-Witten invariants
$r_{g,v}$ has been shown to be captured by these
simpler integral invariants $N_{v,s}$ through
the above map.   In particular the checks made 
for constant maps \ref\fabpan{C. Faber and R. Pandharipande,
``Hodge Integrals and Gromov-Witten theory,'' math. AG/9810173}
and for contribution
of isolated genus $g$ curves to all loops
\ref\pandha{
R. Pandharipande, `` Hodge Integrals and Degenerate Contributions,''
math.AG/9811140.}\
as well as some low genus computations for non-trivial
CY 3-folds
\ref\kkv{S. Katz, A. Klemm
and C. Vafa, ``M-Theory, Topological Strings and
Spinning Black Holes,'' hep-th/9910181.}\ all support the above identification.

Let us illustrate the above results in the case of a simple
non-compact Calabi-Yau threefold, which we will later
use in this paper.  Consider the total space
of the rank 2 vector bundle $O(-1)+O(-1)\rightarrow {\bf P}^1$.
This space has vanishing $c_1$, and is a non-compact
CY 3-fold.  In this case the only BPS state is a membrane
wrapping ${\bf P}^1$ once.  This state has spin $s=0$.  If
we denote the area of ${\bf P}^1$ by $t$, then we have from
\gova\
\eqn\twod{F=\sum_{n>0}{1\over n [2 Sin (n\lambda /2)]^2} e^{-nt}}
For this particular case this has also been derived
using the direct definition of topological strings 
in \fabpan \pandha .

\subsec{Topological Field Theory in 3d:  Chern-Simons Theory}
The 3d topological theory we consider is Chern-Simons theory,
which is given by the Chern-Simons action for a gauge field
$A$:
$$S_{CS}={k\over 4\pi}\int_M Tr[AdA+{2\over 3} A^3]$$
where $M$ is a 3-manifold and
$k$ is an integer which is quantized in order for $exp(iS)$
to be well defined.  As is clear from the definition
of the above action, $S$ does not depend on any metric on $M$
and in this sense the theory is manifestly topological
(i.e. metric independent)\foot{At the quantum level there is a metric
dependence which can be captured by a gravitational Chern-Simons 
term \witcs \ref\axs{S. Axelrod and I. Singer, 
``Chern-Simons Perturbation Theory II,'' J. Diff. Geom. {\bf 39}
(1994) 173.}.}.
  Thus the partition function of Chern-Simons
theory gives rise to topological invariants for 3-manifolds
for each group $G$.  In other words
$$Z_M(G)=exp(-F_M(G))=\int {\cal D}A exp[i S_{CS}]$$
where $A$ is a connection on $M$ for the gauge group $G$
and the above integral is over all inequivalent $G$-connections
on $M$.  The simplest way to compute such invariants is
to use the relation between Hilbert 
space of Chern-Simons theory on a Riemann surface
$\Sigma$ and the chiral blocks of WZW model on $\Sigma$
with group $G$ and level $k$.  For example
the partition function on $S^3$ can be computed by
viewing $S^3$ as a sum of two solid 2-tori, which
are glued along $T^2$ by an order 2 element of $SL(2,{\bf Z})$ 
on $T^2$.  In this way the partition function gets
identified with
$$Z_{S^3}(G,k)= S_{00}(G,k)$$
where $S_{00}=\langle 0|S|0\rangle$ is a particular element of the order
2 operation of $SL(2,{\bf Z})$ on chiral characters, and
is well studied in
the context of WZW models.  In particular for $G=SU(N)$ it is given by:
\eqn\csst{Z_{S^3}(SU(N),k)=exp(-F)=e^{i\pi N(N-1)/8}{1\over (N+k)^{N/2}}
\sqrt{N+k\over N}\prod_{j=1}^{N-1}(2 {\rm sin}{j\pi \over N+k})^{N-j}.}

One can also consider knot invariants:  Consider a knot
$\gamma$ in $M$ and choose a representation $R$
of the group $G$ and consider the character of the holonomoy
of $A$ around the knot $\gamma$, i.e.
$$P[\gamma,R]=Tr_{R} P{\rm exp}(i\int_\gamma A)$$
By the equation of motion for Chern-Simons theory,
which leads to flatness of $A$, we learn that the
above operator only depends on the choice of the knot
type and not the actual knot\foot{In the quantum
theory one also needs to choose a framing for the knot.}.
One then obtains a knot invariant by computing
the correlation function
$$<\prod_iP[\gamma_i,R_i]>=\int {\cal D}A \prod_iP[\gamma_i,R_i]
{\rm exp}(iS_{CS})$$
Again these quantities can be computed by the braiding
properties of chiral blocks in 2 dimensional WZW models
and leads in particular to HOMFLY polynomial invariants for the knots.

\subsec{Topological Field Theories in 4 Dimensions}
If one consider $N=2$ supersymmetric Yang-Mills,
with an unconventional spin assignments, one finds
a topological field theory.  The partition function
is localized on the moduli space of instantons
and the observables of this theory are given by
intersection theory on the moduli space of instantons.
More precisely each $d$-cycle on the four manifold $M$
will lead to a $4-d$ cohomology element on the moduli space
of instantons (obtained by integrating out
$\int {\rm Tr} F\wedge F$ over the corresponding cycle
on the universal moduli space of instantons), and the 
wedging of the cohomology classes gives rise
to the observables in Donaldson theory. This does not
depend on the metric in $M$ (except when
$b_2^+(M)=1$) but will depend on the choice
of smooth structure on $M$.

The computations in this case can be done for many
choices of $M$ by finding
an equivalence of this theory and a simpler abelian theory.
In this case studying the moduli space of non-abelian
instantons gets replaced with the study of an abelian
system known as the Seiberg-Witten equation. The relevant
geometry for the case of $SU(N)$ Yang-Mills is
captured by a certain geometric data related to a
Jacobian variety over an $N-1$ dimensional family of genus $N-1$
Riemann surface, known as Seiberg-Witten geometry \ref\sw{
N. Seiberg and E. Witten, ``Electric-Magnetic Duality, Monopole,
Condensation, and Confinement in N=2 Supersymmetric Yang-Mills
Theory,'' Nucl. Phys. {\bf B426} (1994) 19.}.
For topological field theory aspects 
and how the Seiberg-Witten geometry leads
to computation of the topological correlation functions
see \ref\witsw{E. Witten,
``On S-duality in Abelian Gauge Theory,'' hep-th/9505186.}
\ref\mwi{G. Moore and E. Witten, ``Integration over the u-plane
in Donaldson theory,'' hep-th/9709193.}.

There is another topological theory in 4 dimensions
which has been studied \ref\vwi{C. Vafa and E. Witten,
``A Strong Coupling Test of S-Duality,'' Nucl. Phys.
{\bf B431} (1994) 3.}\
and is related to twisting the maximal supersymmetric
gauge theory in 4 dimensions.  This theory
computes the Euler characteristic of moduli space
of instantons.  In particular for each group $G$ and
each complex parameter $q$
one considers
$$Z_M(G)=q^{-c(M,G)}\sum_k q^{k}\chi ({\cal M}_k)$$
for some universal constant $c$
(depending on $M$ and $G$), where $k$ denotes
the instanton number and $\chi ({\cal M}_k)$ denotes the
euler characteristics (of a suitable resolution and
compactification) of 
${\cal M}_k$, the moduli space of anti-self
dual $G$-connections with instanton number $k$ on $M$.
   Moreover, according to Montonen-Olive
duality conjecture one learns that the above partition function
is expected to be
 modular with respect to some subgroup of $SL(2,{\bf
Z})$ acting in the standard way
on $\tau$ where $q=exp(2\pi i \tau)$. 
For certain $M$ (such as $K3$ ) the above
partition function has been computed and is shown to be
modular in a striking way.
 For recent mathematical discussion on this
see \ref\kap{M. Kapranov, ``The elliptic curve in the
S-duality theory and Eisenstein series for Kac-Moody groups,''
math.AG/0001005.}\ and references therein.

\newsec{Connections between 2d $\leftrightarrow $ 4d TFT's}
There are three different links between 4 dimensional and 2 dimensional
TFT's that I would like to discuss. In all three links the common
theme is that the moduli space of instantons are mapped to
moduli space of holomoprhic curves on appropriate spaces.

\subsec{Topological Reduction of $4d$ to $2d$}
The simplest link between the two theories involves 
studying the 4d TFT on a geometry involving the
product of two Riemann surfaces
$\Sigma_1\times \Sigma_2$, which was studied
in \ref\bjsv{M. Bershadsky, A. Johansen
V. Sadov and C. Vafa, ``Topological Reduction of
4D SYM to 2D Sigma Models,'' Nucl. Phys. {\bf B448} (1995) 166.}.
   In the limit where
$\Sigma_1$ is small compared to $\Sigma_2$ one obtains
an effective theory on $\Sigma_2$ which is the topological
sigma model with target space given by moduli
space of flat connections on $\Sigma_1$, in case
one considers $N=2$ topological field theories
in 4 dimensions or the Hitchin
space associated with $\Sigma_1$ if one
considers $N=4$ topological field theories.  This is natural
to expect because studying light supersymmetric modes
in either case gives rise to the corresponding space
of solutions, which thus behaves from the viewpoint
of the space $\Sigma_2$ as a target space.
In particular the moduli space of 4d instantons
get mapped to moduli space of holomorphic maps
for these target spaces.  Thus quantum cohomology
rings of moduli of flat connections on 
a Riemann surface, which
are encoded in 2d topological correlation functions
 capture the corresponding topological
correlation functions of the 4 dimensional $N=2$ theory.
Similarly in the $N=4$ case the reduction
to 2 dimensions yields a sigma model on the
Hitchin space (which can also be viewed as a Jacobian
variety).  In this context the
 Montonen-Olive duality of $N=4$ theory gets
mapped to mirror symmetry 
of this 2d sigma model (by a fiberwise
application of T-duality to Jacobian fibers).

\subsec{A more subtle 2d $\leftrightarrow$ 4d link}
For the $N=2$ topologically twisted theory, an important
role is played by the Seiberg-Witten geometry, which
is an abelian simplification of the non-abelian 
gauge theory.  This geometry is a quantum deformation
of the classical one, due to pointlike four dimensional
instantons.
  This geometry was first conjectured
based on consistency with various properties
of $N=2$ quantum field theories and its deformation
to $N=1$ quantum field theories with mass gap, where
plausible properties of $N=1$ theories were assumed.

With the recent advances in our understanding of string theory,
the same 4 dimensional gauge theories have been obtained by considering
particular geometries where strings propagate in. This
procedure is known as geometric engineering of QFT's 
(see \ref\kkv{S.
Katz, A. Klemm, and C.
Vafa, ``Geometric Engineering of Quantum Field Theories,''
Nucl. Phys. {\bf B497} (1997) 173.}\ref\kmv{S. Katz, P. Mayr, 
and C. Vafa, ``Mirror Symmetry
and Exact Solution of 4D $N=2$ Gauge Theories," Adv.
Theor. Math. Phys.  {\bf 1} (1998) 53.}\ and references therein).
  These geometries
involve a non-compact Calabi-Yau threefold geometry which is a blow
up of a geometry with some loci of A-D-E singularities
(locally modelled by ${\bf C}^2/G$ where $G$ is a discrete
subgroup of $SU(2)$), giving
rise to the corresponding gauge theory in 4 dimensions. Depending
on the detailed structure of singularities one can obtain
various interesting gauge groups and various matter 
representations.

It turns out that in this description of gauge theory,
the guage theory instantons are mapped to
stringy instantons, which are just worldsheet 
instantons.  Thus being able to compute worldsheet
instantons, i.e. counting of holomorphic
curves in these target geometries, captures the geometry of 4
dimensional gauge theory instantons.
Counting of holomrophic curves is precisely what the (A-model)
topological string computes and thus in this way the 
geometry of vacua of 4 dimensional
gauge theory gets mapped to solving topological amplitudes
in 2d.  This in turn can be done by using (local) mirror
symmetry.  For a physical derivation of mirror
symmetry and some references on this subject
see the recent work \ref\hov{K. Hori and C. Vafa, ``Mirror Symmetry,''
hep-th/0002222.}.
In this way, not only the Seiberg-Witten geometry
has been rederived, but also other geometries
which describe other $N=2$ systems with
various kinds of gauge groups and intricate
matter representations have been obtained \kmv .

\subsec{$N=4$ Yang-Mills on elliptic surfaces and $2d$ topological
theories}

If we consider an $N=4 $ supersymmetric $SU(N)$ topological theory on
an elliptic surface, with base $B$, the stable bundles
get mapped to spectral covers of $B$ on a dual elliptic surface $M$
(where the Kahler class of the elliptic fiber 
is inverted).  This uses the fact that in the limit
of small tori, the stable bundles become flat fiberwise
and flat bundles on tori are related to points on the dual tori.
See \ref\fmw{R. Friedman, J. Morgan, and E. Witten,
``Vector Bundles over Elliptic Fibrations,'' 
alg-geom/9707004.}\ref\bsp{M. Bershadsky, 
A. Johansen, T. Pantev and V. Sadov, ``
On Four Dimensional Compactifications of F-theory,''
Nucl. Phys. {\bf B505} (1997) 165.}\
for a discussion of how this arises.
In particular a rank $N$ stable bundle with instanton number
$k$ gets mapped to a spectral curve which is
a holomorphic curve wrapping the base $N$ times and the elliptic
fiber $k$ times.
Thus the topological $N=4$ amplitude on $M$, denoted
by $Z_M(SU(N))$ which computes
the Euler characteristic of moduli space of $SU(N)$
instantons on $M$ gets mapped to computing
Euler characteristic of moduli space of holomorphic curves (together
with a flat bundle) which in turn is captured by
genus zero topological string amplitudes, and can be
computed using mirror symmetry. This idea has been
implemented in great detail for the case of rational
elliptic surface (also known as ``half $K3$'') \ref\kmv{A. Klemm,
P. Mayr and C. Vafa, ``BPS States of Exceptional Non-critical Strings,''
hep-th/9607139.}\ref\war{J.A. Minahan, D. Nemeschansky and N.P. Warner,
``Partition Functions for BPS States of the Non-critical
$E_8$ String,'' Adv.
Theor. Math. Phys. {\bf 1} (1998) 167.}\ref\nmvw{
J.A. Minahan, D. Nemeschansky, C. Vafa and N.P. Warner,
``E-Strings and $N=4$ Topological Yang-Mills Theories,''
Nucl. Phys. {\bf B527} (1998) 581.}.
The results for the case of rank $2$ 
and its implications for the Euler
characteristic of moduli space of instantons on
rational elliptic surface has been confirmed using rigorous
mathematical methods in \ref\yos{K. Yoshioka, ``
Euler Characteristics of $SU(2)$ instanton moduli spaces on
rational elliptic surfaces,'' Comm. Math. Phys. {\bf 205} (1999)
501.}.

\newsec{Connections between 2d $\leftrightarrow$ 3d TFTs}
Over two decades ago 't Hooft conjectured that $SU(N)$ gauge
theories with
 large $N$  look alot like string theories.  In particular
the partition function for these theories can be organized
in terms of Riemann surfaces where each Riemann surface is weighted
with $N^{\chi}$ where $\chi$ denotes the Euler characteristic
of the Riemann surface.  In particular the low genera dominate
in the large $N$ limit.  The weight factor $N^{\chi}$ follows simply
from the combinatorics of Feynman diagrams, and the Riemann
surface can be identified with the Feynamn diagrams where
the would be holes have been filled.

The main difficulty in the conjecture of `t Hooft is 
to identify precisely which string theory one obtains.
In the past few years for serveral interesting gauge
theories and in particular some in 4 dimensions
the corresponding string theory has been identified
\ref\adscft{O. Aharony, S.S. Gubser, J. Maldacena, H. Ooguri
and Y. Oz,``Large N Field Theories, String Theory and
Gravity,'' Phys. Rept. {\bf 323} (2000) 183.}.  Even though it has
not been possible to actually compute the string theory
amplitudes in these cases, due to the complicated
background strings propagate in, there has been
mounting evidence for the validity of the identification.
One would like to have a similar conjecture in a setup
which is more computable. An ideal setup for this
is topological guage theories, and in particular
the topological Chern-Simons theory.

If we consider $SU(N)$ Chern-Simons theory
on $S^3$ in the limit of large $N$, one could hope to
get a string theory.  It has been conjectured
in \ref\gopv{R. Gopakumar and C. Vafa,
``On the Gauge Theory/Geometry Correspondence,''
hep-th/9811131.}\
that this is indeed the
case.  In particular it has been conjectured
that $SU(N)$ Chern-Simons theory at level $k$
on $S^3$ is equivalent to topological string
with target being a non-compact Calabi-Yau
threefold which is the total space
of $O(-1)+O(-1)\rightarrow {\bf P}^1$, where
the (complexified) size of ${\bf P}^1$ is
given by $t=2\pi iN/(k+N)$ and the string
coupling constant $\lambda ={2\pi i\over N+k}$.  This is a natural
conjecture in the following sense:  The
Chern-Simons theory on $S^3$ can itself be
viewd as an open string theory with target
$T^*S^3$ \ref\wittenoc{E. Witten, ``Chern-Simons Gauge
Theory as a String Theory,'' hep-th/9207094.}.
By open string we mean
considering Riemann surfaces  with boundaries,
where the boundaries are mapped to $S^3$.
The geometry $O(-1)+O(-1)\rightarrow {\bf P}^1$ can
be obtained from the $T^*S^3$ geometry by shrinking
$S^3$ to zero size and blowing ${\bf P}^1$ instead.
This kind of transition is also very similar
to what is observed to happen in the other cases
where large $N$ string theory description was
discovered \adscft .  In fact one can determine \gopv\ the map
of the parameters $t$ and $\lambda $ given above using
this picture (and recalling the metric dependence anomaly in 
Chern-Simons theory).

This conjecture has been checked at the level of the
partition function (which we have briefly reviewed
for both the Chern-Simons theory on $S^3$ and
for $O(-1)+O(-1)\rightarrow {\bf P}^1$ in section 2).
The implications of this conjecture for knot invariants
has been explored in \ref\oov{H. Ooguri and C. Vafa,
``Knot Invariants and Topological Strings,'' hep-th/9912123.}\
and provides a reformulation of knot invariants in terms
of integral invariants which again capture the degeneracy
of spectrum
of (BPS) particles in the corresponding string theory. 
This involves considering a Largrangian submanifold
which intersects $T^*S^3$ along the knot and following it
through the transition to $O(-1)+O(-1)\rightarrow {\bf P}^1$
where it corresponds to a Lagrangian submanifold.  The
corresponding computation on the topological string side
will now involove Riemann surfaces with boundaries, where
the boundary can lie on this Lagrangian submanifold in
$O(-1)+O(-1)\rightarrow {\bf P}^1$.
The results
for the unknot \oov\ as well as the
integrality properties of the torus knots \ref\marila{J.M.F.
Labastida and M. Marino, ``Polynomial Invariants for Torus
Knots,'' hep-th/0004196.}\
 are in perfect agreement with the conjecture.

\newsec{Conclusions}
We have seen some intricate relations among topological
theories in 2, 3 and 4 dimensions and in some ways these
connections parallel the discovery of duality symmetries in
superstring theories (see \ref\vafm{C. Vafa, ``Geometric
Physics,'' Proceedings of ICM-98, hep-th/9810149.}
for a review of some mathematical aspects of string
dualities).
  These topological examples provide a simpler version of superstring
dualities, which one could hope to understand more deeply and which might
provide a hint as to how 
to think about dualities in general.

\vglue 1cm

This research was supported in part by NSF grants PHY-9218167 and
DMS-9709694.

\listrefs

\end